\documentclass[showpacs,showkeys]{revtex4}
\usepackage{amsmath}
\usepackage{graphics}
\def\grad{\vec \nabla }
\begin{document}

\title{Evolution equation for bidirectional  surface waves in a convecting fluid}
\author{ M. C. Depassier}
\affiliation{
 Facultad de F\'\i sica\\
     Universidad Cat\'olica de Chile\\
           Casilla 306, Santiago 22, Chile}

\date{\today}

\begin{abstract}
Surface waves in a heated viscous fluid exhibit a long wave oscillatory instability. The nonlinear evolution of unidirectional waves is known to be described by a modified Korteweg-deVries-Kuramoto-Sivashinsky equation. In the present work we  eliminate the restriction of unidirectional waves and find that the evolution of the wave is governed by a modified Boussinesq system . A perturbed Boussinesq equation of the form 
$y_{tt}-y_{xx} -\epsilon^2(y_{xxtt} + (y^2)_{xx })+ \epsilon^3( y_{xxt}+y_{xxxxt} + (y^2)_{xxt}) =0 $ which includes instability  and dissipation can be derived from this system.
\end{abstract}


\maketitle

\section{Introduction}
\label{intro}

Long wave oscillatory instabilities arise in different contexts. They have been found in the study of unstable drift waves in plasmas \cite{Cohen76}, fluid flow along an inclined plane \cite{Benney66,Topper78}, convection in fluids with a free surface \cite{BD87,BD89,AD90,GV92}, the Eckhaus instability of traveling waves \cite{Janiaud92}  and, more recently, in solar dynamo waves \cite{Mason05}. 
The nonlinear evolution of the instability is generically described by a Korteweg-deVries-Kuramoto-Sivashinsky (KdV-KS) equation
$$
u_t +  u u_x + \delta  u_{xxx} + r u_{xx} + u_{xxxx}=0,
$$ 
in some cases with different or additional nonlinearities \cite{AD90,Janiaud92,Mason05}. The solutions to this equation have been studied extensively focusing  mainly on the formation and interaction of solitary pulses \cite{Kawahara83,Kawahara88,Balmforth97}.  It exhibits parameter regimes where different behavior is found, numerical studies have shown the appearance of modulated traveling waves, period doubling and chaos \cite{Alfaro94,Balmforth97}. Exact solutions have been constructed by different methods \cite{Lou91,Porubov96,Parkes02,Kudryashov04}. The limits of large and small dispersion $\delta$ have been studied rigourously \cite{Biagoni96,Iosevich02}. 

The derivation of this equation in the physical situations mentioned above involves the assumption of unidirectional waves. This is  natural  in the situations where a preferred direction exists; the electron diamagnetic direction in the case of plasma waves,  towards the equator in the case of solar dynamo waves,  downhill for flow along the inclined plane. However in the case of surface waves in a convecting fluid this assumption is not natural. The assumption of unidirectional waves for surface waves was abandoned in \cite{Kraenkel94, Nepom94} and a modified Boussinesq system of equations derived. 

 The purpose of this article is to reexamine this problem to find the nonlinear evolution equation of the instability without the assumption of unidirectional waves, which will be the natural extension of the KdV-KS equation to bidirectional waves. A higher order Boussinesq system is derived which, following the usual procedure for water waves \cite{RSJohnson},  can be reduced to a modified Boussinesq equation including instability and dissipation.  Our main interest is to find the  form of this equation, rather than focusing on a particular physical problem.
The equation we find has been introduced as one of two possible heuristic extensions of the Boussinesq equation to dissipative systems \cite{Christov94}, to our knowledge it has not been derived in a systematic way for any physical problem. In this work we show that the modified Boussinesq equation is derived systematically for systems with long wave oscillatory instabilities, in a manner analogous as the Boussinesq equation is derived for surface waves in an ideal fluid.
 For this reason we take as a model the simplest case in which this instability can be found,  more precisely, we consider surface waves in a convecting fluid neglecting surface tension.  

\section{Formulation of the problem}

We consider a layer of fluid which, at rest, lies between $z=0$
and $z=d$. Upon it acts a gravitational field $\vec g =- g\hat z$. The
fluid is described by the Boussinesq equations
\begin{eqnarray}
 \nabla \cdot \vec v &=& 0
  \\
 \rho _0 \frac{d\vec v}{dt} &=& -\grad p + \mu \nabla ^2 \vec v
+ \vec g \rho
 \\
\frac{dT}{dt} &=& \kappa \nabla^2 T 
\end{eqnarray}
where the density $\rho$ depends linearly on the temperature
\begin{equation}
\rho = \rho_0 [1-\alpha (T-T_0)].
\end{equation}
Here $\frac{d }{dt} = \frac{\partial}{\partial t} + \vec v \cdot
\grad$ is  the convective derivative; p, T,  and $\vec v$
denote  the pressure, temperature and fluid velocity
respectively.  The quantities $\rho_0$ and $T_0$ are reference
values.  The fluid properties, namely its viscosity $\mu$, thermal
diffusivity $\kappa$, and coefficient of thermal expansion $\alpha$
are constant. Furthermore we restrict ourselves to two dimensional
motion  so that $\vec v =(u,0,w)$.

The fluid is bounded above by a free surface on which the heat flux
is fixed and  upon it a constant pressure $p_a$ is exerted. 
Below, it is bounded by a plane stress-free surface which is
maintained at constant temperature. As the fluid moves the free
surface is
deformed,  we shall denote  its position by $z=d+\eta (x,t)$. The
boundary  conditions on the upper surface are
\begin{eqnarray}
& & \eta_t + u \eta_x = w  
\\
& & p-p_a - \frac{2 \mu}{N^2} [w_z +u_x \eta_x^2 -\eta_x (u_z+w_x)] = 0  
 \\
& & \mu (1-\eta_x^2)(u_z+w_x)+2 \mu \eta_x(w_z-u_x)=0  
 \\
 & & \hat n \cdot \grad T = -F/k  
 \end{eqnarray}

on $z=d+\eta$.  
 Here, subscripts denote derivatives,   
$N=(1+\eta_x^2)^{1/2},$
$\hat n = (-\eta_x,0,1)/N$ is the unit normal to the free surface,
 $F$ is the prescribed normal heat flux, and  $k$ is the
thermal conductivity.

Denoting by $T_b$ the fixed temperature of the lower surface, the
 boundary conditions on the lower surface $z=0$
are
\begin{equation}
w = u_z =0,  \qquad   T= T_b 
\end{equation}

The static solution to these equations is given by
$
T_s = -F(z-d)/k + T_0, \, 
\rho_s = \rho_0 [1+(\alpha F/k)(z-d)],$ and $
p_s = p_a -g \rho_0 [(z-d)+(\alpha F/2k)(z-d)^2].$
We have chosen the reference temperature $T_0$ as the value of the
static temperature on the upper surface. The temperature on the lower
surface is then $T_b = T_0 + Fd/k$.
Equations (1 - 9) constitute the problem to be solved.
We shall
adopt $d$ as unit of length, $d^2/{\kappa}$ as unit of time,
$\rho_o d^3$ as unit of mass, and $Fd/k$
as unit of temperature. 
 Then  there are three  dimensionless parameters 
involved in the problem, the Prandtl number $\sigma =\mu / {\rho_o
\kappa}$, 
the Rayleigh
number $R=\rho_o g \alpha F d^4  / {k \kappa \mu}$, and  the Galileo number
$G= g d^3 \rho_0^2 / \mu^2$.

The linear stability theory has been studied elsewhere \cite{BD87,BD89}. The result of interest in the present context is the existence of an oscillatory instability at vanishing wave number $a_c=0$ with   critical Rayleigh number $R_c=30$. The  frequency along the marginal curve is given by   $\omega = a \sigma \sqrt{G} + \ldots$ which vanishes  at criticality. In dimensional quantities this frequency is that of long surface waves in an ideal fluid. In this problem the damping due to viscosity is compensated by heating. 

\section{Small amplitude nonlinear expansion}

To study the evolution of the oscillatory instability slightly above onset we let
$$
R= R_c + \epsilon^2 R_2
$$
and, since the instability occurs at vanishing wavenumber and the frequency along the marginal curve is of the order of the wave number, we introduce scaled   time and horizontal coordinates
$$
\xi = \epsilon x, \qquad \tau = \epsilon t.
$$
Next we look for a perturbative solution expanding the dependent variables as
\begin{eqnarray*}
& & p = p_s(z) + \epsilon^2 ( p_0 + \epsilon p_1 + \epsilon^2 p_2 + \ldots)\\
& & T = T_s(z) + \epsilon^3 ( \theta_0 + \epsilon \theta_1 + \theta^2 p_2 + \ldots)\\
& & u = \epsilon^2 ( u_0 + \epsilon u_1 + \epsilon^2 u_2 + \ldots)\\
& & w = \epsilon^3 ( w_0 + \epsilon w_1 + \epsilon^2 w_2 + \ldots)\\
& & \eta = \epsilon^2 (\eta_0 + \epsilon \eta_1 + \epsilon^2 \eta_2 +\ldots),
\end{eqnarray*}
and proceed to solve at each order. Our main interest is to obtain a modified Boussinesq system including instability and dissipation. This requires that the calculation be carried out to order $\epsilon^3$. For the sake of simplicity we choose the Prandtl number and the Galileo number equal to 1. This choice, and the neglect of surface tension effects,  does not introduce any qualitative effects. The details are given in the Appendix, here we quote the relevant results.

In leading order the horizontal velocity is given by 
$$
u_0 = f(\xi,\tau),
$$
where $f(\xi,\tau)$ is an arbitrary function to be determined. At each higher order a new arbitrary function appears in the solution for the horizontal speed. We will find a coupled system of equations for the surface height and horizontal velocity which are a modified Boussinesq system. The first equation of this system is obtained from the kinematic boundary condition for the free surface, the second equation  from the solvability condition of the equations (1-9).  

The kinematical boundary condition is, at each order,
\begin{eqnarray}
& & \eta_{0\tau} + f_{\xi} = 0 \\
& & \eta_{1\tau} + g_{\xi} = 0 \\
& & \eta_{2\tau} + h_{\xi} + \alpha f_{\xi\xi\xi} + ( f \eta_0)_{\xi} = 0 \\
& & \eta_{3\tau} + j_{\xi} + \alpha g_{\xi\xi\xi} - \beta f_{\xi\xi\xi \tau} + (f \eta_1 + g \eta_0)_{\xi} = 0.
\end{eqnarray}

The functions $g, h$ and $j$ are  arbitrary functions of $\xi$ and $\tau$.
The coefficients $\alpha$ and $\beta$ are positive numbers, $\alpha = 71/168$, $\beta = 5149/30240$. The second equation needed to close the system comes, as mentioned above, from a solvability condition. At each order we find,
\begin{eqnarray}
& & \eta_{0 \xi} + f_{\tau} = 0 \\
& & \eta_{1 \xi} + g_{\tau} = 0 \\
& & \eta_{2 \xi} + h_{\tau}-\gamma f_{\xi\xi\tau} + f f_{\xi} + 30  \eta_0 \eta_{0 \xi}   = 0 \\
& & \eta_{3 \xi} + j_{\tau}-\gamma g_{\xi\xi\tau} + r f_{\xi\xi} + \tilde\mu f_{\xi\xi\xi\xi}  + \tilde\nu f_{\xi\xi\tau \tau} + \\
& & \qquad \qquad \qquad \qquad \qquad 16 (\eta_0 f_{\xi})_{\xi} + 30  (\eta_0 \eta_1)_{\xi} + (f g)_{\xi}  = 0. \nonumber
\end{eqnarray}
In these equations $r= 2 R_2/15$ is the measure of instability, the rest are numeric parameters. For our choice $\sigma= G =1$, their values are 
$\tilde\mu = 478/699$, $\tilde\nu = 4897/10080$ and $\gamma = 257/168$.
Finally we construct a single system, correct to order $\epsilon^3$  for the surface elevation $\eta$ and arbitrary function $F = f + \epsilon g + \epsilon^2 h + \epsilon^3 j$ by simply adding the equations above.  In doing so some ambiguity arises in the choice of terms of order $\epsilon^2$ and $\epsilon^3$, where it is possible to replace $\eta_{\xi} = -F_{\tau} + O(\epsilon^2)$ or to replace $F_{\xi} = -\eta_{\tau} + O(\epsilon^2)$.  In the theory of ideal water waves this leads to the existence of different families of Boussinesq systems \cite{Bona2002}. At this stage we use this freedom  to minimize the type of terms. With this criterion, the kinematic condition, correct to order $\epsilon^3$ becomes
\begin{equation}
\eta_{\tau} + F_{\xi} +\alpha \epsilon^2 F_{\xi\xi\xi} - \beta \epsilon^3 
F_{\xi\xi\xi \tau} + \epsilon^2 (F \eta)_{\xi} = 0.
\label{primera}
\end{equation}
The solvability condition correct to order $\epsilon^3$ is
\begin{equation}
 \eta_{ \xi} + F_{\tau}-\epsilon^2 \gamma F_{\xi\xi\tau} + \epsilon^3 r F_{\xi\xi}-\epsilon^3 (\tilde\mu   + \tilde\nu ) \eta_{\xi\xi\xi \tau} + 
  \epsilon^2  F F_{\xi}   + 30  \epsilon^2 \eta \eta_{\xi} + 16 \epsilon^3 (\eta F_{\xi})_{\xi} = 0. 
\label{segunda}
\end{equation}

This is a modified Boussinesq system including instability and dissipation in the linear terms of order $\epsilon^3$. It describes the propagation of bidirectional waves.  In the following section,  we reduce this system to a single equation for the surface elevation, obtaining a modified Boussinesq equation.

\section{Evolution equation for the free surface}

If we search for unidirectional waves the  system (10-17) leads to a modified KS-KdV equation \cite{AD90}
\begin{equation}
\eta_t + \delta_1 \eta \eta_x + \delta_2 \eta_{xxx} +
 \epsilon (r \eta_{xx} +\delta_3
\eta_{xxxx} +\delta_4 (\eta \eta_x)_x) = 0
\label{unidir}
\end{equation}
Our purpose here is to obtain the evolution equation for the surface without this restriction. In this process some ambiguity arises as explained in the previous section. In reducing to a single equation we adopt the criterion of choosing the form that gives the simplest physically meaningful linear theory. 
\subsection{Linear Theory}

Keeping linear terms in equations (\ref{primera},\ref{segunda}), we find that the surface displacement satisfies the equation
$$
\eta_{\tau\tau} - \eta_{\xi\xi} - \epsilon^2 A \eta_{\xi\xi\tau\tau} + \epsilon^3 ( r \eta_{\xi\xi\tau} + B \eta_{\xi\xi\xi\xi\tau} ) =0,
$$
where $A = \alpha + \gamma$, $\mu_2 = \tilde \mu + \tilde \nu + \beta$.  Searching for plane wave solutions ${\rm e}^{s \tau} {\rm e}^{ i k \xi}$, we obtain the characteristic polynomial
$$
s^2 ( 1 + \epsilon^2 A k^2) - \epsilon^3 (r k^2 - B k^4) s + k^2 = 0.
$$
The growth rate is given by
$$
\lambda = \Re(s) = \frac{1}{2} \frac{{\rm \epsilon}^3 (r k^2 - B k^4)}{1 + \epsilon^2 A k^2}
$$
and the frequency  is given by 
$$
\omega^2 = \Im(s)^2 = \frac{ k^2 - \frac{1}{4} \epsilon^6 (r k^2 - B k^4)^2}{1 + \epsilon^2 A k^2}.
$$
 On the marginal curve $\lambda = 0$,  $r = \mu_2 k^2$.  Going back to unscaled variables, we recover the expansion for the Rayleigh number $R = R_c + a^2 B$ valid at small wavenumbers. Small wavenumbers are unstable, large wavenumbers are stable as it occurs in the KdV-KS equation. 

\subsection{Nonlinear equation}

Taking cross derivatives of equations  (\ref{primera},\ref{segunda}) to obtain a wave equation for the surface elevation we obtain
\begin{equation}
\eta_{\tau\tau} - \eta_{\xi\xi} - \epsilon^2 A \eta_{\xi\xi\tau\tau} + \epsilon^3 ( r \eta_{\xi\xi\tau} + B \eta_{\xi\xi\xi\xi\tau} ) + \epsilon^2 [(F \eta)_{\xi\tau} - \frac{1}{2}(F^2)_{\xi\xi}] - 15 \epsilon^2 (\eta^2)_{\xi\xi} + 8 \epsilon^3 (\eta^2)_{\xi\xi\tau}=0.
\label{paso1}
\end{equation}
We recognize the order $\epsilon^2$ containing $F$ as already appearing in the derivation of the standard Boussinesq equation, therefore we only outline the procedure \cite{RSJohnson} to obtain the modified Boussinesq equation .
We write
$$
[(F\eta)_{\tau} -  F F_{\xi}]_{\xi} = - \left( \frac{1}{2} \eta^2 + F^2\right)_{\xi \xi} + O (\epsilon^2),
$$
and 
$$
(F^2)_{\xi\xi} = 2 \left(\eta_{\tau}^2 - F \eta_{\tau\xi}\right) + O(\epsilon^2).
$$
Assuming that $F$ vanishes at $-\infty $ we may write 
$$
F = - \int_{-\infty}^\xi \eta_{\tau} d\xi' + O(\epsilon^2)
$$
Replacing these expressions in (\ref{paso1}) we obtain
$$
\eta_{\tau\tau} - \eta_{\xi\xi} - \epsilon^2 A \eta_{\xi\xi\tau\tau} + \epsilon^3 ( r \eta_{\xi\xi\tau} + B \eta_{\xi\xi\xi\xi\tau} )  - 2 \epsilon^2 \left( \eta_{\tau}^2 + \eta_{\tau\xi} \int_{-\infty}^\xi \eta_{\tau} d\xi'\right)
- \frac{31}{2} \epsilon^2 (\eta^2)_{\xi\xi} + 8 \epsilon^3 (\eta^2)_{\xi\xi\tau}=0.
$$
Changing  to new variables (Y, T),
$$
Y = \xi + \epsilon^2 \int_{-\infty}^\xi \eta_{\tau} d\xi',\qquad \qquad T = \tau
$$
and defining
$$
N= \eta - \epsilon^2 \eta^2
$$
 we obtain our main result, a single evolution equation for bidirectional waves including instability and dissipation,
\begin{equation}
N_{TT} - N_{YY} -\epsilon^2 A N_{YYTT} -\frac{33}{2} \epsilon^2 (N^2)_{YY} + \epsilon^3 (r N_{YYT} + B N_{YYYYT}) + 8 \epsilon^3 (N^2)_{YYT} =0.
\label{final}
\end{equation}
Up to order $\epsilon^2$ we recognize the improved Boussinesq equation. Instability and dissipation are included in the linear terms of order $\epsilon^3$. The nonlinearity of order $\epsilon^3$ is an additional nonlinearity which is characteristic of this problem (see Eq. (20)), and which is also found in the study of the Eckhaus instability \cite{Janiaud92}. 

\section{Summary}

We have studied the nonlinear development of a long wave oscillatory instability of surface waves.   When the assumption of unidirectional waves is made the surface evolution is determined by a modified KdV-KS equation. In the present study we allow bidirectional propagation of waves and are led to a modified Boussinesq system, where instability and dissipation are present at high order.  A single equation is obtained for the evolution of the surface following the method used in ideal surface waves. In the case of ideal surface waves one is led to a Boussinesq equation , here we obtain a modified Boussinesq equation including instability and dissipation.

Previous studies have dealt with the damped Boussinesq equation ($ r <0$, B=0) where stable structures,  solitons and periodic solutions,   must be sustained by external forcing \cite{Arevalo2002,Maccari2004}. In the present case $r>0$ and it represents the destabilizing term. The equation we find has been proposed as one of two possible heuristic generalizations of the KdV-KS equation to bidirectional waves. Here we show that it is indeed the adequate bidirectional generalization of the KdV-KS equation by deriving it from a physical problem, just as the Boussinesq equation is derived for surface waves in an ideal fluid.   Numerical and analytical results for this modified Boussinesq equation (\ref{final}) will be the subject of future work.

\section*{Acknowledgements}
Partial support  from Fondecyt  grant 1060627 is acknowledged.

\appendix*
\section{Perturbation expansion}

In this appendix the equations to be solved at each order and their
solution are given. Terms that vanish have been omitted.

In leading order the equation to be solved are
\begin{eqnarray*}
u_{0zz} &=& 0 \\
w_{0z} &=& -u_{0 \xi}  \\
    p_{0z} &=& 0 \\
\theta_{0zz} &=& -w_0 
\end{eqnarray*}
subject to 
\begin{eqnarray*}
& &w_0(0)= u_{0z}(0) = \theta_0(0) = 0 \\ 
& &u_{0z}(1) = \theta_{0z}(1) = 0 \\
& &p_0(1) =  \eta_0 \\
& &  \eta_{0 \tau} = w_0(1).
\end{eqnarray*}
The solution is given by
\begin{eqnarray*}
& &u_0 = f(\xi ,\tau ),  \qquad w_0 = - f_\xi z \\
& &\theta_0= f_\xi T_0(z), \qquad p_0 =  \eta_0 
\end{eqnarray*}
where we have defined $T_0(z) = (z^3-3z)/6$. The kinematical boundary condition is
\begin{equation}
 \eta_{ 0 \tau} + f_{\xi} =0.
\end{equation}

At order $\epsilon$ the equations to be solved are
\begin{eqnarray*}
u_{1zz} &=& p_{0\xi} + u_{0\tau} \\
w_{1z} &=& - u_{1\xi} \\
p_{1 z} &=& 30 \, \theta_0 \\
\theta_{1 zz} &=& \theta_{0 \tau} - w_1
\end{eqnarray*}
subject to
\begin{eqnarray*}
& &w_1(0)= u_{1z}(0) = \theta_1(0) = 0 \\
& &u_{1z}(1) = \theta_{1z}(1) = 0 \\
& &p_1(1) =  \eta_1 + 2 w_{0z}(1)\\
& & \eta_{1 \tau} = w_1(1).
\end{eqnarray*}
The solvability condition $\int_0^1 u_{1zz} dz = 0$ implies
\begin{equation}
\eta_{0 \xi} + f_\tau =0.
\end{equation}
The solution at this order is
\begin{eqnarray*}
& &u_1 = g(\xi ,\tau ),  \qquad w_1 = - g_\xi z \\
& &\theta_1= g_\xi T_0(z) + f_{\xi \tau} T_1(z), \qquad p_1 =  \eta_1 + 30 f_\xi P_1(z) - 2 f_\xi 
\end{eqnarray*}
where we have defined $T_1(z) = (25 z - 10 z^3 + z^5)/120$ and $P_1(z) = T_{1z}(z)$. The kinematical boundary condition becomes
\begin{equation}
 \eta_{ 1 \tau} + g_{\xi} =0.
\end{equation}
At order $\epsilon^2$ the equations to be solved are
\begin{eqnarray*}
u_{2zz} &=& p_{1\xi} + u_{1\tau}- u_{0\xi\xi} \\
w_{2z} &=& - u_{2\xi} \\
p_{2 z} &=& 30 \, \theta_1 + w_{1 z z} - w_{0 \tau}\\
\theta_{2 z z} &=& \theta_{1 \tau} - w_2 - \theta_{0 \xi \xi}
\end{eqnarray*}
subject to
\begin{eqnarray*}
& &w_2(0)= u_{2z}(0) = \theta_2(0) = 0 \\
& &u_{2z}(1) = -w_{0\xi}(1), \qquad \theta_{2z}(1) =  -\eta_0 \theta_{0 z z}(1) \\
& &p_2(1) =  \eta_2 + 2 w_{1z}(1) + 15 \eta_0^2 - \eta_0 p_{0 z}(1)\\
& & \eta_{2 \tau} = w_2(1) - (u_0 \eta_0)_\xi.
\end{eqnarray*}
The solvability condition $\int_0^1 u_{2zz} dz = u_{2z}(1)$ implies
\begin{equation}
\eta_{1 \xi} + g_\tau =0,
\end{equation}
and the solution to this order is
\begin{eqnarray*}
& &u_2 = h(\xi ,\tau ) + f_{\xi\xi} U(z),  \qquad w_2 = - h_\xi z  - f_{\xi\xi\xi} W(z)\\
& &\theta_2= h_\xi T_0(z) + g_{\xi \tau} T_1(z)+ f_{\xi \tau \tau} T_2(z) + f_{\xi\xi\xi} T_3(z) - z \eta_0 f_\xi, \\
& & p_2 =  \eta_2 + 15 \eta_0^2 +  30 g_\xi P_1(z) + f_{\xi \tau} P_2(z)  - 2 g_\xi .
\end{eqnarray*}
where we have defined $U(z) = z^2(39-15 z^2 + z^4)/24$, $W(z) = (91 z^3 - 21 z^5 + z^7)/168$, $P_2(z)= (25 - 30 z^2 + 5 z^4)/120$,  $T_2(z) = (25 z - 10 z^3 + z^5)/120$, $T_3(z) = (5 z^9-180 z^7+1134 z^5+5040 z^3-19575 z)/60480$.  The kinematical boundary condition becomes
\begin{equation}
 \eta_{ 2 \tau}  + h_{\xi} + \alpha f_{\xi\xi\xi} + ( f \eta_0)_{\xi} = 0.
\end{equation}
At order $\epsilon^3$ we only need to calculate the solvability condition for $u_3$ and the kinematical boundary condition. The equation for the horizontal velocity is
$$
u_{3zz} = p_{2\xi} + u_{2\tau}- u_{1\xi\xi} + u_0 u_{0\xi}
$$ subject to
$$ u_{3 z}(0) = 0, \qquad  u_{3z}(1)  = - w_{1\xi}(1).$$
The solvability condition yields 
\begin{equation}
\eta_{2 \xi} + h_{\tau}-\gamma f_{\xi\xi \tau} + f f_{\xi} + 30  \eta_0 \eta_{0 \xi}   = 0,
\end{equation}
and the solution for the horizontal velocity is
\[
u_3 = j(\xi ,\tau ) + g_{\xi\xi} U(z) + f_{\xi\xi \tau} U_3(z) + (\eta_{2\xi} + h_\tau + f f_\xi + 30 \eta_0 \eta_{0\xi})z^2/2.
\]
In the  expression above $U_3(z) = (-1022 z^2 + 294 z^4 - 28 z^6 + z^8)/672.$
 
The kinematical boundary condition at this order 
\[
\eta_{3 \tau} = w_3(1) + (u_0 \eta_1)_\xi + (u_1 \eta_0)_{\xi}
\]
yields
\begin{equation}
\eta_{3 \tau} + j_{\xi} + \alpha g_{\xi\xi\xi} - \beta f_{\xi\xi\xi \tau} + (f \eta_1 + g \eta_0)_{\xi} = 0,
\end{equation}
The pressure satisfies the equation
\[p_{3 z} = 30 \, \theta_2 + w_{2 z z} - w_{1 \tau} + w_{0\xi\xi} + R_2 \theta_0
\]
with
\[ p_3(1) = \eta_3 + 30 \eta_0\eta_1 + 2 w_{2z}(1) -\eta_0 p_{1z}(1).\]
Its explicit solution is not indispensable to continue the calculation.
Finally, the last equation needed is obtained from the solvability condition of the horizontal velocity at order $\epsilon^4$. We have
\[
u_{4zz} = p_{3\xi} + u_{3 \tau} - u_{2\xi\xi} + (u_0 u_1)_\xi, \]
with
\[ u_{4z}(0)=0, \qquad u_{4z}(1) = -\eta_0 u_{2zz}(1) - w_{2\xi}(1) -\eta_0 w_{0 \xi z}(1) - 2 \eta_{0\xi} (w_{0z}(1)- u_{0\xi}(1)). \]
The solvability condition leads to Eq. (17).

\end{document}